\documentclass[angeo, manuscript]{copernicus}
\newcommand{\Deg}{\degree}

\nolinenumbers
\begin{document}

\title{Launching mass from the Moon helped by lunar gravity anomalies}


\Author[FMI,APT][pekka.janhunen@fmi.fi]{Pekka}{Janhunen} 

\affil[FMI]{Finnish Meteorological Institute, Erik Palm\'enin aukio 1, FI-00560 Helsinki, Finland}
\affil[APT]{Aurora Propulsion Technologies, Otakaari 5, FI-02150 Espoo, Finland}




\runningtitle{Launching mass from the Moon}

\runningauthor{Janhunen}

\received{}
\pubdiscuss{} 
\revised{}
\accepted{}
\published{}


\firstpage{1}

\maketitle

\begin{abstract}
  Normally a passive object launched from the Moon at less than the
  escape velocity orbits the Moon once and then crashes back to the
  launch site. We show that thanks to lunar gravity anomalies, for
  specific launch sites and directions, a passive projectile can remain
  in lunar orbit for up to 9 Earth days. We find that such sites exist
  at least on the lunar equator for prograde equatorial orbit
  launches. Three of the sites are located on the lunar nearside.  We envision
  that this can be used to lift material from the Moon at low cost
  because it gives prolonged opportunities for an active spacecraft to
  catch the projectile.  Passive projectiles can be made entirely from
  lunar material so that a stream of Earth-imported parts is not needed. To reduce
  the mass and cost of the launcher, the projectile mass can be scaled
  down with a corresponding increase in the launch frequency.  The
  projectile launcher itself can be a coilgun, railgun, superconducting
  quenchgun, sling or any other device that can give a projectile an
  orbital speed of about 1.7 km/s.
\end{abstract}


\introduction  

There is a need to get material, especially propellant, into cislunar
orbits at a lower cost than launching it from Earth. Human
exploration of Moon and Mars requires chemical propellant, especially
liquid oxygen (LOX) because it dominates the mass budget relative to
the fuel. LOX could be sourced from the Moon by extracting O$_2$ from
the regolith \citep{LomaxEtAl2020}. There is also an emerging
technology to use O$_2$ as an electric propulsion propellant
\citep{AndreussiEtAl2017}. The separation of oxygen from the regolith
leaves a silicon and metal -rich residue that could, with further
industrial steps, be processed into structural materials like
aluminium and steel. Already 50 years ago, \citet{ONeill1974} proposed
to launch lunar material with an electromagnetic mass driver (coilgun)
to build large rotating settlements -- later called O'Neill cylinders
-- for people to live in under artificial gravity. The plan was to fund the activity by
building large solar power satellites \citep{Glaser1968} by the same
orbital construction infrastructure \citep{SP428} and selling their
energy to the grid. Recently, new solar power satellites geometries
were discovered that avoid single failure points in the form of large
rotary joints \citep{Cash2019}, and this has spurred renewed interest
in solar power satellites. This plethora of applications
(exploration, electric propulsion, solar power satellites, habitats)
suggests that there is a need to find low-cost ways of lifting lunar material to
orbit.

Lifting mass from the Moon can be done with reusable chemical rockets,
but for good economy the propellants needed for the lifting should be
sourced from the Moon. As mentioned above, LOX can be sourced from the
lunar regolith, but the production of fuel such as H$_2$ or CH$_4$
requires volatile elements that are scarce except in the permanently
shadowed craters near the poles \citep{SowersAndDreyer2019}. Operating
in the permanently shadowed craters presents challenges. The total
inventory of water-equivalent hydrogen on the Moon is roughly
estimated to be $\sim 10^{11}$ kg \citep{HurleyEtAl2016}. Although
large, the amount is nevertheless limited when considering widespread
sustainable use. Thus, because of the mining complexities and the
somewhat limited amount of fuel resources, it would be desirable to be
able to lift lunar material by something else than a chemical rocket.

To address this, electric-powered projectile launcher
concepts for lifting lunar material have been considered, such as
coilgun \citep{ONeill1974,SP428,EhresmannEtAl2017}, railgun \citep{McNabEtAl2022},
superconducting quenchgun \citep{NottkeAndBilby1990} and sling
\citep{BakerAndZubrin1990,Landis2005}. After being launched, the
projectile must be captured by a satellite that takes it to a central
processing facility. The central facility might be located e.g.~at the
Earth-Moon Lagrange L5 point. Alternatively, the projectile might
guide itself to the central facility directly, but then the projectile
would need an active propulsion system. For the sake of economy and
simplicity, it is desirable that the projectile is made entirely of
lunar resources with no Earth-imported parts. Thus a scheme that works with passive projectiles is desirable.

If a passive projectile is launched with less than the lunar escape
velocity, it normally falls back to the Moon after completing one
lap. Capturing it by a satellite during the single lap would be
challenging. If the launch speed matches the escape velocity, then the
projectile can escape the Moon's gravitational sphere of influence. In
theory it could reach e.g.~a station at L5. However, tracking the
passive projectile over the vast distance (equal to the Earth-Moon
distance of 384400 km) and the trajectory uncertainty would increase
toward the end. Any failed attempts would create untracked
cislunar debris objects.

The Moon's gravity field is complicated with many local anomalies in
multiple scales.  The lunar gravity anomalies are often considered
problematic, because they would typically make low lunar orbiting
satellites crash the surface after some time unless propulsive control is
applied. The crashing occurs because
the anomalies tend to change the eccentricity of the satellite's
orbit, so that an initially circular orbit turns into a more elliptical
one whose perilune is lower than the original orbit.

In this paper we consider the following idea.  Since the gravity
anomalies can increase the eccentricity of an initially circular orbit
to make it crash the surface, time symmetry of the equation of motion
suggests that a reverse process should also exist under some
circumstances. We try and find lunar launch locations where the
gravity anomalies conspire in such a way that the passive projectile
orbits the Moon several times before crashing, thus giving more time
for an active lunar orbiting satellite to capture it and carry it to a
destination such as L5, typically using low-thrust electric propulsion.

\section{Model}

We use the order-1200 spherical harmonic model of the lunar gravity
field derived from Gravity Recovery and Interior Laboratory (GRAIL)
mission \citep{LemoineEtAl2014,GoossensEtAl2016}. The gravity model is
given in the body-fixed Principal Axis (PA) coordinate system.  On the
lunar surface, the PA coordinate system differs by about 1 km from the
Mean Earth (ME) coordinate system used in lunar maps.  For the
purposes of this paper, the difference between the PA and ME
coordinates systems is sufficiently small to be safely neglected.

We calculate the trajectory of a small point mass in the spherical
harmonic model gravity field using the leapfrog time integrator. The
integration timestep is 1 s. For detecting collisions with the lunar
surface, we use a model of lunar topography provided with the GRAIL
gravity model dataset. It has been derived from Lunar Orbiter Laser
Altimeter (LOLA) instrument \citep{SmithEtAl2010} onboard Lunar
Reconnaisance Orbiter (LRO). The Moon's rotation is included in the
model. The program is written in C++ and uses the GNU Scientific
Library functions for computing the associated Legendre polynomials
that arise when evaluating the spherical harmonic expansion.

\section{Results}

The intention is to launch a passive projectile from the lunar surface
so that it be grabbed by an active lunar orbiting satellite. Normally
a passive object launched from a celestial body will crash near the
launch point after completing one lap unless, and in that case the
grabbing satellite would have only a single opportunity to try and
catch it. However, our hope is to find some launch points for which
the lunar gravity anomalies would increase the perilune altitude and
thus conspire to our favour, enabling a passive projectile to orbit
the Moon several times before crashing and thus providing more
realistic opportunities for a satellite to catch it.

We restrict ourselves to equatorial launch locations and equatorial
orbits because then the catching satellite passes over the launch
point at every orbit, thus having a launch window every orbit, i.e.,
approximately every two hours. We also consider only the nearside of
the Moon (longitude between 90{\Deg}W and 90{\Deg}E) because then a
direct radio link with Earth is possible, and we consider only
prograde orbit i.e.~eastward launches.

\begin{figure}[t]
\includegraphics[width=8.3cm]{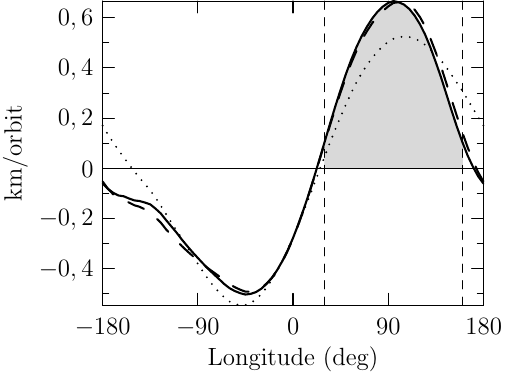}
\caption{Change rate of perilune altitude for 10x300 km equatorial
  orbit, as function of perilune longitude. Solid line is full order-1200
  gravity model, dashed line is truncated order-60 and dotted line
  order-20. Grey area marks the longitude range where the perilune increases by at least 0.1 km per orbit.}
\label{fig:scanfig}
\end{figure}

To get an overview of how the gravity anomalies affect a projectile
launched from the lunar surface, we imagine that we launch the
projectile from a virtual 10 km high tower (above the mean lunar
radius) and ignore possible collisions with the lunar surface. As the
apolune altitude we use 300 km so that the initial orbit is slightly
eccentric. We perform the calculation with the GRAIL gravity anomalies
so that we can learn how the longitude of the launch site affects the
evolution of the orbit. We are interested in finding longitude ranges
where the perilune altitude increases during the (first) orbit. Figure
\ref{fig:scanfig} shows the result. The longitude range where the
per-orbit perilune increase is positive and at least 0.1 km in
magnitude is marked in grey.

Figure \ref{fig:scanfig} shows that longitudes between
30{\Deg}E and 160{\Deg}E are potentially feasible for launching a passive
projectile that orbits the Moon at least several times before
crashing, although the projectile -- being passive -- does not perform
any orbit circularisation burn to raise its perilune altitude. So for
the lunar nearside, the interesting longitude range is 30{\Deg}E--90{\Deg}E.

In Fig.~\ref{fig:scanfig}, the solid line shows the full order-1200
gravity model result while the dashed and dotted lines show truncated
orders 60 and 20, respectively. The order-60 result is close to the
full result. The order-20 result differs significantly although it
still retains the top-level qualitative features. This suggest that
the full order-1200 result is robust because it can be truncated down
to 60 without changing the conclusions.

The feasible launch site are equatorial mountaintops within
30{\Deg}E--90{\Deg}E eastern longitude, that do not have nearby higher
mountains on the eastern side. Also the mountaintop must connect by
rover-accessible paths to large nearby areas that also include mare
regions, in order to have the option of launching also mare-derived
minerals. We conservatively define rover-accessible to mean at most a
10-degree slope. Although many rovers can climb also slopes steeper
than 10\Deg, the potential presence of rocks (for which we do not have
data in this paper) might make the path harder, and therefore we
prefer to exercise caution regarding the slope.

\begin{figure}[t]
\includegraphics[width=8.3cm]{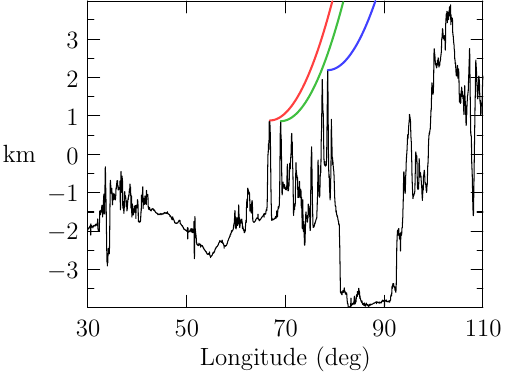}
\caption{Equatorial lunar profile and 300 km apolune trajectories
  from Site 1 (red), 2 (green) and 3 (blue).}
\label{fig:proffig}
\end{figure}

With these criteria, we have identified three potential launch sites (Fig.~\ref{fig:proffig}),
which we now present in more detail in the following subsections. For
each site we simulate a launch to the apolune altitude of 300 km
(above the mean equatorial lunar surface) and do full collision
detection with respect to the lunar topography at every timestep until
the projectile crashes. Along the way we plot the evolution of the
perilune altitude, defined by computing a sliding minimum over the orbital
period of the instantaneous altitude relative to the local topography.

\subsection{Site 1: 0{\Deg}N, 66.852{\Deg}E}

\begin{figure}[t]
\includegraphics[width=8.3cm]{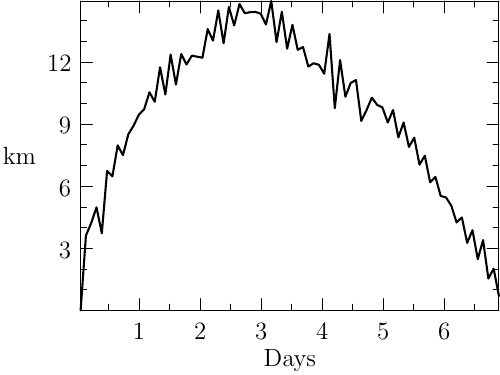}
\caption{Time evolution of perilune altitude for Site 1.}
\label{fig:site66}
\end{figure}

The equatorial site with longitude of 66.852{\Deg}E is between
Maclaurin crater and Mare Spumans and south from Pomortsev crater. The
site has a low-slope (max 10\Deg) access to both highlands and mares.
Figure \ref{fig:site66} shows the time evolution of the perilune.  The
projectile crashes after 7.0 Earth days (167 hours) and the maximum
perilune reached is over 14 km.

\subsection{Site 2: 0{\Deg}N, 69.0547{\Deg}E}

\begin{figure}[t]
\includegraphics[width=8.3cm]{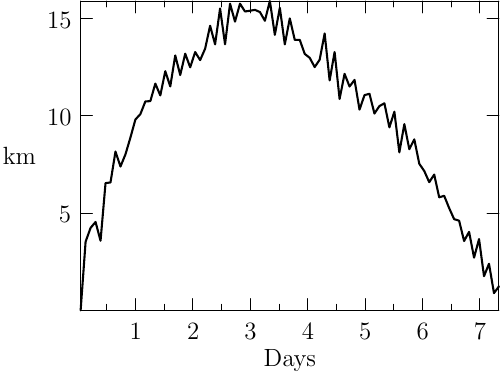}
\caption{Time evolution of perilune altitude for Site 2.}
\label{fig:site69}
\end{figure}

The longitude 69.0547{\Deg}E site is on the eastern rim of Maclaurin X
crater.  There is a $\sim$15 km long few hundred metres narrow
low-slope (max 10\Deg) access path from the south, where it connects
to large areas including mares. Figure \ref{fig:site69} shows the time
evolution.  The projectile crashes after 7.5 Earth days (180 hours)
and reaches maximum perilune of about 15 km. Immediately after launch,
the projectile's minimum clearance from the nearby high ground is less
than for Sites 1 and 3 (Fig.~\ref{fig:proffig}).

\subsection{Site 3: 0{\Deg}N, 78.57734{\Deg}E}

The longitude 78.57734{\Deg}E site is a rim between Jenkins and
Schubert J craters.  Along the rim a low-slope access path exists,
connecting to wide regions into south and probably into north as
well. The projectile crashes after 9.54 Earth days (229 hours) and
reaches maximum perilune of over 20 km.

Site 3 seems the most promising because it has the highest perilune
and the longest orbital lifetime, but the other two sites are feasible
too.

\begin{figure}[t]
\includegraphics[width=8.3cm]{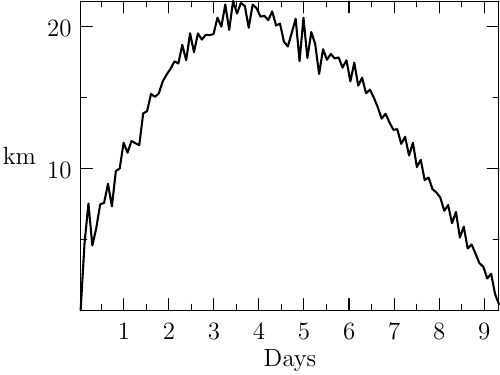}
\caption{Time evolution of perilune altitude for Site 3.}
\label{fig:site78}
\end{figure}

\begin{figure*}[t]
\includegraphics[width=12cm]{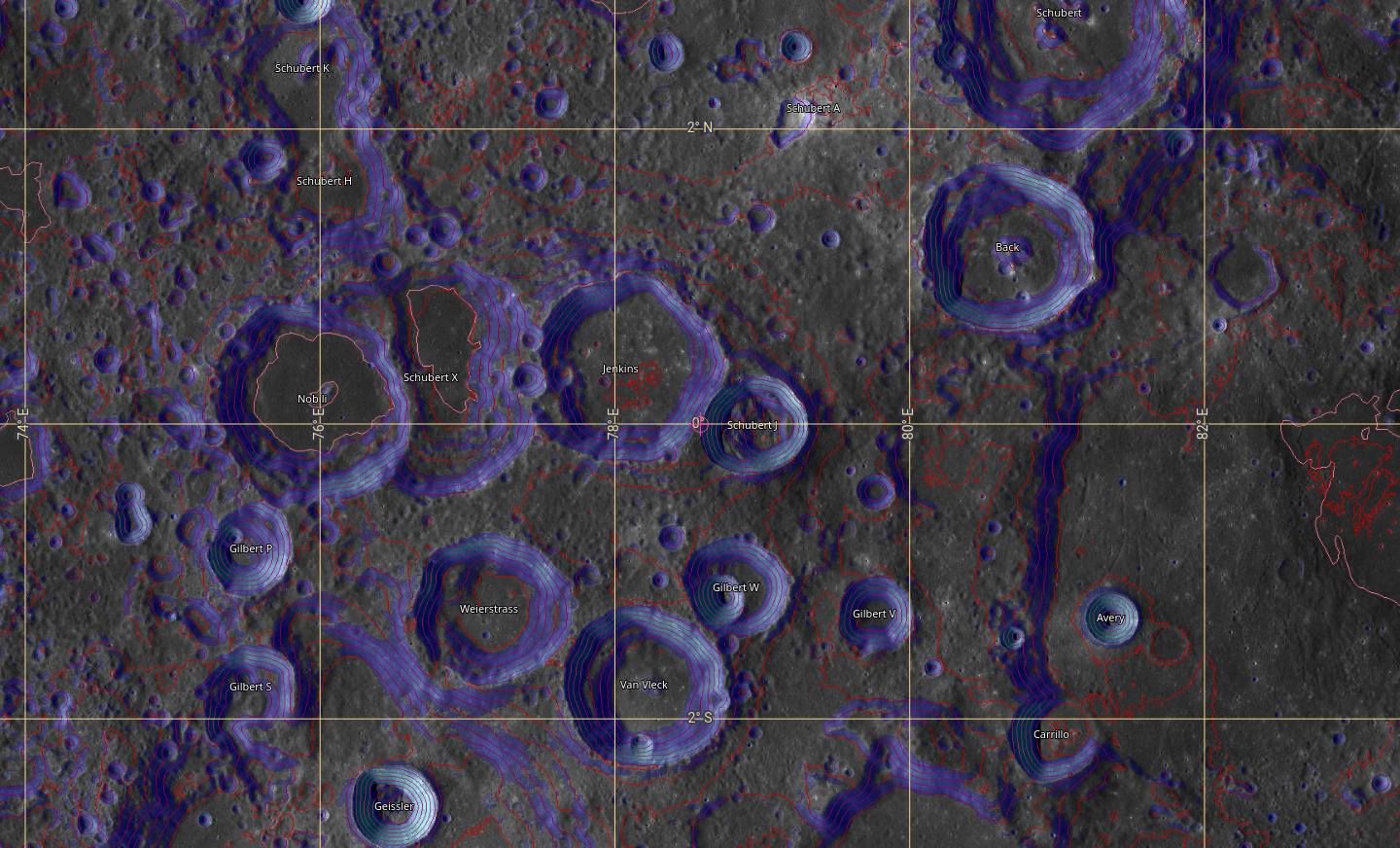}
\caption{Context of Site 3 on the rim between Jenkins and
  Schubert J craters (pink circle with crosshairs at the centre). Areas with slope steeper than 10\Deg are
  shaded. Terrain height is shown by red contours with 500 m
  spacing. The 2\Deg grid spacing corresponds to 60.7 km.}
\label{fig:site78screendump}
\end{figure*}

Figure \ref{fig:site78screendump} shows a map of Site 3,
produced by the Quickmap tool provided by the Lunar Reconnaisance Orbiter
Camera (LROC) team.


\begin{figure}[t]
\includegraphics[width=8.3cm]{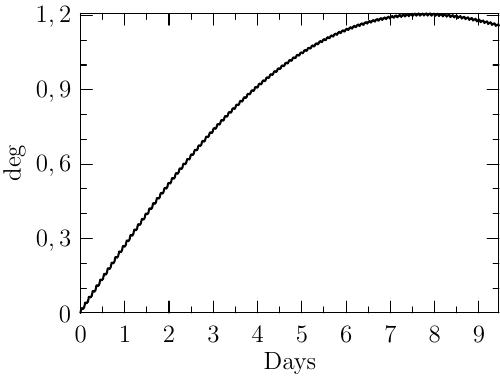}
\caption{Time evolution of orbital inclination for Site 3.}
\label{fig:site78incl}
\end{figure}

The orbital inclination evolves slightly, reaching a maximum of
1.2\Deg after 8 Earth days in orbit (Fig.~\ref{fig:site78incl}). One
might want to compensate for it by launching slightly off-east,
so that at the time of capture the orbit is more closely equatorial.

\section{Discussion}

We checked numerically that inclusion of Earth's tidal effects does
not affect the results appreciably. Sun's tidal effects are even
smaller.


We saw above (Fig.~\ref{fig:scanfig}) that truncating the originally
order-1200 spherical harmonic expansion at order-60 still produces
almost the same results, whereas truncating it further down to order-20
changes the results significantly. This is consistent with the fact
that the average orbital altitude in our case was about 150 km (the
average between nearly zero perilune and 300 km apolune). The
circumference of the lunar equator divided by 60 gives 182 km
horizontal resolution for the order-60 spherical harmonic
expansion. Spatial structure in the surface gravity field whose scale
size is much less than the orbital altitude is damped out upon
reaching the orbital altitude. Although the GRAIL mission managed to
reveal the gravity anomalies up to spherical harmonic order-1200 from
orbit, it was possible only because it had precise orbit determination
and other sophisticated features. Thus the differences between the
solid, dashed and dotted curves in Fig.~\ref{fig:scanfig} are
understandable.

Lunar regolith can be melted e.g.~in a solar oven to produce
glass-like material \citep{MeurisseEtAl2018}. Although presently most
produced glassy samples from lunar regolith simulant have demonstrated
relatively low compressive strengths of less than 5 MPa
\citep{MeurisseEtAl2018}, efforts are underway to improve the
strength.  Therefore, producing projectiles that tolerate the
acceleration of the launcher system will probably be feasible in the
future.  The acceleration produced by the launcher system depends on
how it was designed.  Also, for a given level of acceleration, the
required compressive strength scales linearly with the projectile's
size, so that a small projectile has an easier time withstanding the
launch than a large one.  Although details depend on the chosen
launcher technology, in general it is likely that the projectile
launching can be arranged so that the projectile only needs to
withstand compression, which is easier for a glassy material to
tolerate than tension.

Lifting mass from the Moon to low lunar orbit by an electric launcher
is energetically cheap. For 1740 m/s speed the specific electric
energy is only 1.44 MJ/kg. (The 1740 m/s is the shooting speed to an
equatorial prograde orbit with 300 km apolune altitude. The small 4.5 m/s helping
effect due to lunar rotation has been taken into account.) This is
more than an order of magnitude less than the energy required to
electrolyse water to produce hydrogen-oxygen rocket propellant for
performing a corresponding chemical rocket launch. The reason is that
when using a rocket, a lot of the energy goes into accelerating the
exhaust gases rather than the payload.

Besides on the mountaintop itself, the launch site could also be
located on the eastward-facing slope of the same mountain, as long as
it is sufficiently high up so that the projectile does not risk
hitting any neighbouring high ground in the east. Depending on what kind
of projectile launcher one uses, it may be more advantageous to use
the mountaintop itself or its eastward-facing slope.

The active satellite that captures the projectile from low lunar orbit
could use low-thrust propulsion (electric propulsion), because it has
several Earth days to accomplish the capture operation. Different
capture strategies can be envisioned. One way is to perform a
traditional rendezvous where the relative velocity between the
capturing satellite and the projectile is gradually zeroed before the
physical capture. Another way is to use some device such as a bag that
allows capture with nonzero relative speed.

One possible scenario is that one collects the projectiles
at the Earth-Moon Lagrange L5 point where there is a factory that
extracts oxygen from the projectile e.g.~by using the process
described by \citet{LomaxEtAl2020}. The low-thrust delta-v difference
between the low lunar orbit and the L5 is 700--1700 m/s. The lower
limit 700 m/s corresponds to thrusting at perilune so that the Oberth
effect is used maximally and the upper limit 1700 m/s corresponds to
continuous thrusting and spiralling-out orbit. At 2000 s specific
impulse which is typical to electric propulsion, 3.6--9\% of the mass must be
propellant. If the electric propulsion thruster uses oxygen as the propellant
\citep{AndreussiEtAl2017}, then the transfer chain can be self-contained regarding propellant.

\conclusions  

We showed that by launching material in the form of passive
projectiles from certain mountaintops on the lunar equator (three of
which are located on the lunar nearside), the projectile can remain in
equatorial low lunar orbit for up to 9 Earth days before crashing to
the surface. During this time it can be captured by a satellite more
easily than in the normal case where the projectile returns after one lap.
The phenomenon is due to the lunar gravity anomalies. Because the projectiles are passive, they
can be made entirely from lunar material so that no importing of
material from Earth is needed. They can be scaled down in mass to
reduce the mass and cost of the launcher system.  The actual
projectile launcher options include coilgun, railgun, superconducting
quenchgun, sling or any other device that can give a projectile an
orbital speed of 1740 m/s.

If successfully developed, the method may enable low-cost and scalable
lifting of lunar material for obtaining propellant and construction
materials in cislunar space. Lunar polar volatile mining is not
necessarily required because the lifting does not need chemical
propulsion.

In the future, one should study further the rendezvous and capture
strategies, the trade-offs related to the projectile mass, and the
advantages and disadvantages of the various projectile launcher
alternatives. If needed, it would be straightforward to generalise the
present study to include farside launch sites, nonzero inclinations,
polar orbits, different apolune altitudes and retrograde orbits.









\noappendix       







\authorcontribution{PJ developed the model code, performed the simulations and prepared the manuscript} 

\competinginterests{The authors declare that they have no conflict of interest.} 


\begin{acknowledgements}
The results presented have been achieved under the framework of the
Finnish Centre of Excellence in Research of Sustainable Space (FORESAIL), Academy
of Finland grant number 352849.
\end{acknowledgements}

\end{document}